\documentclass[12pt]{article}

\usepackage{epsfig}
\usepackage{amsmath}
\usepackage{amssymb}

\begin{document}

\begin{titlepage}
\begin{flushright}
HU-EP-00/22 \\
\today
\end{flushright}

\vspace{1cm}
\begin{center}
\baselineskip25pt
{\Large\bf A Small Cosmological Constant and Backreaction of
           Non-Finetuned Parameters}
\end{center}
\vspace{1cm}
\begin{center}
\baselineskip12pt
{Axel Krause\footnote{Now at University of Maryland, Department
of Physics, College Park, Maryland 20742, USA;\\
E-mail: {\tt krause@physics.umd.edu}}}
\vspace{.3truecm}

\vspace{1cm}

{\em
\centerline{Humboldt-Universit\"at zu Berlin,}
\vspace{2mm}
\centerline{Institut f\"ur Physik, D-10115 Berlin, Germany}}

\vspace{0.3cm}
\end{center}
\vspace*{\fill}

\begin{abstract}
We include the backreaction on the warped geometry induced by
non-finetuned parameters in a two domain-wall set-up to obtain
an exponentially small Cosmological Constant $\Lambda_4$. The
mechanism to suppress the Cosmological Constant involves one
classical fine-tuning as compared to an infinity of finetunings
at the quantum level in standard D=4 field theory.
\end{abstract}


\vspace*{\fill}

\end{titlepage}

Starting with heterotic M-theory \cite{HW1}, \cite{HW2} as the
prototypical fundamental brane world scenario where Grand
Unification (GUT) becomes also a unification with higher
dimensional gravity a lot of activity has been invested to
explore this and other brane worlds. One of the
model-independent phenomenological successes of heterotic
M-theory is that with the input of the GUT scale $M_{GUT}$ and
the value of the GUT gauge coupling $\alpha_{GUT}$ it predicts
a lower bound on the effective four-dimensional Newton's
Constant $G_N$ which coincides strikingly with its measured
value \cite{GN}. The geometries of these brane worlds are
typically described by warped geometries which allowed to solve
the hierarchy problem in a novel way which makes essential use
of the warped extra dimensions \cite{RS1}. It is thus natural
to investigate the usefulness of warped geometries also for the
largest hierarchy problem, the cosmological constant problem
\cite{CC}. While it was proposed in \cite{KS} that warped
geometries could be used to explain a vanishing cosmological
constant, it was proposed in \cite{AK3} that they might lead to
a mechanism for obtaining a small cosmological constant of
order $\Lambda_4 \simeq (\text{meV})^4$. While these are
mechanisms which assume a field-theory framework to deal with
the issue of the cosmological constant one should keep in mind
that in string-theory there are examples of three-dimensional
vacua with negative and zero cosmological constants which are
connected with each other through T-Duality and therefore
obscure the precise low-energy meaning of the vacuum energy
\cite{CCT}. There might also be a completely new understanding
of the vacuum energy if M-theory turns out to be a theory of
only a finite \cite{Banks} but huge amount of discrete
chain-like degrees of freedom as suggested in
\cite{KCT} based on microscopic black hole entropy derivations.
The non-local chains and the associated discreteness of
spacetime should shed some new light on how quantum field
theory has to be modified in the UV.

We will in this paper restrict ourselves to the traditional
field-theory framework and explore further the mechanism for
obtaining a small cosmological constant proposed in \cite{AK3}.
The mechanism used a five-dimensional set-up consisting of two
four-dimensional positive-tension $T > 0$ domain-walls (there is
no need for either the bulk or the walls to be supersymmetric)
separated by a distance $2l=M_{GUT}^{-1}$ ($M_{GUT}$ = Grand
Unification scale) along the fifth noncompact dimension.
Together with bulk gravity and a non-positive bulk cosmological
constant $\Lambda(x^5)\le 0$ the set-up is described by the
action
\begin{equation}
  S=-\int d^5x \left( \sqrt{-G}\left[ M^3R(G)+\Lambda(x^5) \right]
                     +\sqrt{-g^{(4)}}T\left[ \delta(x^5+l)+\delta(x^5-l)
                                     \right]
               \right) \; .
  \label{WallAction}
\end{equation}
Neither of the walls is conceived as hidden but instead they are
both thought of as being the origin of the Standard Model
fields. E.g.~by a string-embedding of the set-up and the
realisation of the domain-walls as two stacks of D3-branes, one
can think of the Standard-Model gauge group SU(3) as arising
from one stack and the $\text{SU(2)}\times\text{U(1)}$ from the
other \cite{AK3}. For finetuned parameters this set-up leads to
a warped geometry containing a flat 4-dimensional spacetime
section
\begin{alignat}{3}
  ds^2 &= e^{-A(x^5)}\eta_{\mu\nu}dx^\mu dx^\nu + (dx^5)^2 \;\; ; \;\;
                            \mu,\nu=1,\hdots,4 \notag \\
  A(x^5) &= \frac{k}{2}
            \left( |x^5+l|+|x^5-l| \right) \;\; , \;\;
  k = \sqrt{-\Lambda_e/3M^3}
  \label{FlatSolution}
\end{alignat}
with the bulk cosmological constant $\Lambda(x^5)$ and
wall-tension $T$ given by
\begin{equation}
  \Lambda(x^5) =
            \left\{ \begin{array}{cc}
      \Lambda_e   &, \;\;|x^5| > l \\
      \Lambda_e/4 &, \;\;|x^5| = l \\
          0       &, \;\;|x^5| < l
                    \end{array}
            \right.
  \; , \qquad
  T=\sqrt{-3M^3 \Lambda_e} \; .
  \label{Tension}
\end{equation}
such that the effective four-dimensional Cosmological Constant
$\Lambda_4$ vanishes. In this paper, we want to determine the
full backreaction of non-finetuned parameters on the warped
geometry and demonstrate that with one fine-tuning the resulting
effective $\Lambda_4$ comes out exponentially suppressed thus
freeing the effective four-dimensional theory from the need to
correct the cosmological constant order by order in
perturbation theory. The suppression-length will turn out to be
given by the distance between both walls and will have to be
chosen by the inverse of the GUT scale as explained in detail
in \cite{AK3}. To this aim, we have to determine the resulting
5-dimensional geometry for general non-positive $\Lambda \le 0$
and positive $T>0$.

Let us start with a $D$-dimensional warped geometry
\begin{equation}
  ds^2=G_{MN}dx^M dx^N=f(x^D)g_{\mu\nu}(x^\rho,x^D)dx^\mu dx^\nu + (dx^D)^2 \; ,
\end{equation}
where $\mu,\nu,\rho=1,\hdots,D-1$ and $f(x^D)$ denotes the
warp-factor. While the warp-factor in this Ansatz is supposed
to be a differentiable function of $x^D$, the dependence of the
4-dimensional metric $g_{\mu\nu}(x^\rho,x^D)$ on $x^D$ is
assumed to be piecewise constant with possible jumps only
occuring at the position of the two walls. The freedom to allow
for such jumps at the walls comes from the observation that one
should expect the walls to have some finite thickness. Most
conservatively this is estimated to be of the order of the
$D$-dimensional Planck-length. In a fundamental brane-world
theory like heterotic M-theory \cite{HW1} into which the present
two wall set-up could be embedded, the Planck-length is
actually larger than naively expected -- it is of the same
order as the inverse of the Grand Unification scale \cite{BD}
and indeed is believed to correspond to the width of the walls
(orbifold fixed planes) \cite{HW2}. Therefore, the width seems
large enough not to neglect {\it a priori} variations in the
$(D-1)$-dimensional curvature over the width. If the variation
occurs rapid enough then we can model the situation in the
approximation of infinitesimal width by a jump in the
$(D-1)$-dimensional geometry. Indeed, it has been shown in
\cite{BV} that fundamental brane world theories very similar to
\cite{HW2} which originate from M-theory and use M9 branes for
the walls can only be constructed if one allows for a stepwise
constant $(D-1)$-dimensional curvature paired with a stepwise
constant $D$-dimensional cosmological constant with jumps at
the brane positions (where $D=11$ for the case of M-theory).
Similarly, when we turn to $D=5$ we will later distinguish
between a 4-dimensional cosmological constant $\Lambda_4(x^5)$
which is piecewise constant and valid on 4-dimensional sections
of the 5-dimensional spacetime and $\Lambda_4$ which appears
after integrating over $x^5$ in the effective 4-dimensional
action and is of course $x^5$ independent. As long as the
4-dimensional observer cannot resolve the fifth dimension the
parameters of the effective 4-dimensional theory are obtained
by integrating out the fifth dimension.

The induced metric on a $(D-1)$-dimensional section defined by
$x^D=$ const, will be denoted by $g^{(D-1)}_{\mu\nu}(x^\rho,x^D)
=f(x^D)g_{\mu\nu}(x^\rho,x^D)$. Our aim is to solve the Einstein
equation piecewise in $x^D$ (such that $g_{\mu\nu}(x^\rho,x^D)$
is constant with respect to $x^D$ on every such piece and $x^D$)
to first determine the lower-dimensional $\Lambda_4(x^5)$ for
the case $D=5$. Therefore, we decompose the $D$-dimensional
Ricci-tensor $R_{MN}$ into its $\mu$ and $D$ components
\begin{alignat}{3}
  R_{\mu\nu}(G) &= R_{\mu\nu}(g)+\frac{1}{4}g_{\mu\nu}
            \left( 2f^{\prime\prime}+(D-3)f\left[(\ln f)^\prime\right]^2
            \right) \notag \\
  R_{\mu D}(G) &= 0 \\
  R_{DD}(G) &= \frac{1}{4}(D-1)\left( 2\frac{f^{\prime\prime}}{f}
                                    - \left[(\ln f)^\prime\right]^2
                               \right) \; . \notag
\end{alignat}
This allows to decompose the $D$-dimensional Einstein-tensor
$E_{MN}(G)=R_{MN}-\frac{1}{2}R(G) G_{MN}$ as
\begin{alignat}{3}
  E_{\mu\nu}(G) &= E_{\mu\nu}(g)+g_{\mu\nu}\frac{(D-2)}{2}
                   \left[ \left(1-\frac{(D-1)}{4}\right)
                          f\left[(\ln f)^\prime\right]^2
                         -f^{\prime\prime}
                   \right] \notag \\
  E_{\mu D}(G) &= 0 \\
  E_{DD}(G) &= -\frac{1}{2f}R(g)
               -\frac{(D-1)(D-2)}{8}
                \left[(\ln f)^\prime\right]^2 \; . \notag
\end{alignat}

Let us now restrict ourselves to the $D=5$ case, where the
expressions simplify to
\begin{alignat}{3}
  E_{\mu\nu}(G) &= E_{\mu\nu}(g)-\frac{3}{2}g_{\mu\nu}f^{\prime\prime}
   \notag \\
  E_{\mu 5}(G) &= 0
   \label{Ein2} \\
  E_{55}(G) &= -\frac{1}{2f}R(g)
               -\frac{3}{2}\left[(\ln f)^\prime\right]^2
    \; . \notag
\end{alignat}
For the action (\ref{WallAction}) specifying the set-up, the
gravitational sources consist of a non-positive bulk
cosmological constant $\Lambda(x^5)
 \le
0$ and walls with tension $T$ placed at $x^5=l$ and $x^5=-l$,
which amounts
 to
the following energy-momentum tensor
\begin{equation}
  T_{MN} = -\Lambda(x^5) G_{MN} - T\left[\delta(x^5+l)+\delta(x^5-l)\right]
            g^{(4)}_{\mu\nu}\delta^\mu_M\delta^\nu_N \; .
\end{equation}
Decomposing the 5-dimensional Einstein-equation,
$E_{MN}(G)=-T_{MN}/(2M^3)$, with the help of (\ref{Ein2}) into
its $\mu$ and 5 components, we receive from the $\mu\nu$ part
the 4-dimensional Einstein-equation
\begin{equation}
  E_{\mu\nu}(g)=\left[ \frac{3}{2}f^{\prime\prime}+\frac{f}{2M^3}
                       \left[ \Lambda(x^5)+T\delta(x^5+l)+T\delta(x^5-l)
                       \right]
                \right] g_{\mu\nu} \; .
  \label{FourEinstein}
\end{equation}
From the $55$ part follows an expression for the 4-dimensional
curvature
 scalar
\begin{equation}
  R(g)=-f\left[3\left[(\ln f)^\prime\right]^2+\frac{\Lambda(x^5)}{M^3}
         \right] \; ,
  \label{FourScalar}
\end{equation}
whereas the $\mu 5$ part is satisfied trivially.

Contraction of $E_{\mu\nu}(g)$ with $g^{\mu\nu}$ gives
$E^\mu_{\phantom{\mu}\mu}(g)=\frac{3-D}{2} R(g)\rightarrow
-R(g)$ and
 therefore
leads to the following consistency equation among
(\ref{FourEinstein}) and (\ref{FourScalar})
\begin{equation}
    2\frac{f^{\prime\prime}}{f}-\left[(\ln f)^\prime\right]^2
  = -\frac{1}{3M^3}\left[ \Lambda(x^5)+2T\delta(x^5+l)+2T\delta(x^5-l)
                   \right] \; .
    \label{Consistency}
\end{equation}

It is evident that the right-hand-sides of (\ref{FourEinstein})
and (\ref{FourScalar}) must be piecewise constant with respect
to $x^5$, since both left-hand-sides are at least piecewise
independent of $x^5$. This is a consequence of the simple
warp-factor Ansatz. It means that the 4-dimensional sections
$\Sigma_4$, defined by $x^5=const$, must be spacetimes of
constant curvature. For $R(g)<0$ we have de Sitter
($\text{dS}_4$) and for $R(g)>0$ Anti-de Sitter ($\text{AdS}_4$)
spacetime. Since this already determines the solution to the
Einstein
 equation
up to a scalar quantity -- the curvature -- the equations
(\ref{FourEinstein}),(\ref{FourScalar}),(\ref{Consistency})
become linear dependent and it suffices to solve only two of
them.

When we foliate the 5-dimensional spacetime into sections
$\Sigma_4$, we see that the Einstein-equations
(\ref{FourEinstein}),(\ref{FourScalar}) also follow from the
4-dimensional action on $\Sigma_4$
\begin{equation}
   S_{D=4}(x^5) =-\int_{\Sigma_4} d^4x \sqrt{-g}\left( M^2_{\text{eff}} R(g)
                                          +\Lambda_4(x^5)
                                    \right)
\end{equation}
if we make the following identifications\footnote{The
4-dimensional sections exhibit
                 \begin{equation*}
                    E_{\mu\nu}(g) = \frac{\Lambda_4(x^5)}{2M_{\text{eff}}^2}
                                    g_{\mu\nu} \; , \qquad
                    R(g) = -2\frac{\Lambda_4(x^5)}{M_{\text{eff}}^2} \; ,
                 \end{equation*}
with $\text{dS}_4:\; R(g)<0,\Lambda_4(x^5)>0$ and
$\text{AdS}_4:\; R(g)>0,\Lambda_4(x^5)<0$.}
\begin{alignat}{3}
    \frac{3}{2}f^{\prime\prime}+\frac{f}{2M^3}
    \left[ \Lambda(x^5)+T\delta(x^5+l)+T\delta(x^5-l) \right]
 &= \frac{\Lambda_4(x^5)}{2M_{\text{eff}}^2}
     \label{FourScalarEquiv1} \\
    -f\left[3\left[(\ln f)^\prime\right]^2+\frac{\Lambda(x^5)}{M^3}
      \right]
 &= -\frac{2\Lambda_4(x^5)}{M_{\text{eff}}^2} \; .
     \label{FourScalarEquiv2}
\end{alignat}
Notice that the dependence of $S_{D=4}(x^5)$ and of the
cosmological constant $\Lambda_4(x^5)$ on sections with respect
to $x^5$ is a piecewise constancy. Here $M_{\text{eff}}$ is the
effective Planck-scale, as obtained by integrating the
5-dimensional action (\ref{WallAction}) over $x^5$
\begin{equation}
  M_{\text{eff}}^2 = M^3 \int dx^5 f(x^5) \; .
\end{equation}
The Einstein equations (\ref{FourEinstein}),(\ref{FourScalar})
now become replaced by
(\ref{FourScalarEquiv1}),(\ref{FourScalarEquiv2}).

To recognize the relation between the piecewise constant
cosmological constant $\Lambda_4(x^5)$ on sections $\Sigma_4$
and the final effective $\Lambda_4$ obtained by integrating out
the fifth dimension of (\ref{WallAction}), we note that
$\Lambda_4$ is given by \cite{AK3}
\begin{equation}
  \Lambda_4 = \int dx^5 f^2
              \bigg(M^3 \big[ [(\ln f)^\prime]^2
                                +4\frac{f^{\prime\prime}}{f}
                         \big]
                     +   \big[ \Lambda(x^5)+T\delta(x^5+l)+T\delta(x^5-l)
                         \big]
              \bigg) \; .
  \label{DefinitionLambda}
\end{equation}
Using (\ref{FourScalarEquiv1}) for the second term in square
brackets, we obtain the simple and expected relationship
\begin{equation}
  \Lambda_4 = f^\prime f |^{x_R^5}_{x_L^5} + \langle\Lambda_4(x^5)\rangle
                                                                    \; ,
  \label{DeterminingLambda}
\end{equation}
where $x^5_R,x^5_L$ denote the right and left boundary of the
$x^5$ integration region and the mean is weighted with the
profile of the
 warp-factor
\begin{equation}
         \langle\Lambda_4(x^5)\rangle
  \equiv \frac{\int dx^5 f \Lambda_4(x^5)}{\int dx^5 f} \; .
\end{equation}
Since we will see that the total derivative contribution
$f^\prime f |^{x_R^5}_{x_L^5}$ will vanish in our case of
interest, we learn that the 4-dimensional effective action
$S_{D=4}$ is related to the sectionwise action by taking the
mean, $S_{D=4}=\langle S_{D=4}(x^5) \rangle$.

Since only two of the equations
(\ref{Consistency}),(\ref{FourScalarEquiv1}),(\ref{FourScalarEquiv2})
are independent, it is most convenient to choose
(\ref{Consistency}) to determine the warp-factor in terms of
the fundamental ``input'' parameters $\Lambda(x^5),M$ and $T$.
In a further step, we will then obtain $\Lambda_4(x^5)$ from
(\ref{FourScalarEquiv2}). Expressing the warp-factor through
$f=e^{-A(x^5)}$ and denoting $Y(x^5)=A^\prime(x^5)$, we
 can
write (\ref{Consistency}) as
\begin{equation}
  -2Y^\prime+Y^2+\frac{\Lambda(x^5)}{3M^3}
 =-\frac{2T}{3M^3}\left[ \delta(x^5+l)+\delta(x^5-l) \right] \; ,
    \label{YEquation}
\end{equation}
With the signature-function defined by $\text{sign}(x)=-1$ if
$x\le 0$ and $\text{sign}(x)=1$ if $x>0$, the solution to this
differential equation is given by
\begin{equation}
  Y(x^5)= -\frac{k}{2}
      \left(\text{sign}(x^5+l)+\text{sign}(x^5-l)\right)
      \coth\left(\frac{k}{4}
                 \left[ |x^5+l| + |x^5-l| - 2a \right]
           \right)
        \label{YSol}
\end{equation}
together with the following $\Lambda(x^5)$ profile with
arbitrary but non-positive constant $\Lambda_e \le 0$
\begin{equation}
  \Lambda(x^5)  = \left\{ \begin{array}{cc}
                  \Lambda_e &,\;\; |x^5| > l \\
                  \Lambda_e/4 \le 0&,\;\; |x^5| = l \\
                      0  &,\; |x^5| < l
                          \end{array}
                   \right.
   \label{CosmStep}
\end{equation}
and the wall-tension
\begin{equation}
  \frac{T}{3M^3} = k\coth\left(\frac{k}{2}(a-l)\right) \; .
   \label{IntConstTension}
\end{equation}

Here, as in the introduction, $k=\sqrt{-\Lambda_e/3M^3}$ and
$a$ is an integration constant. The last relation which
determines $a$ through the bulk cosmological constant
$\Lambda_e$ and the wall-tension $T$ has been gained
 by
satisfying the boundary conditions at the wall-locations, which
are encoded
 in
the $\delta$-function terms in (\ref{YEquation}). A matching of
the $\delta$-function terms arising from $Y^\prime$ with those
proportional to
 $T$
leads to (\ref{IntConstTension}). The symmetry of the set-up --
caused by the equality of both wall-tensions
 --
forces the bulk cosmological constant between them to be zero. A
 non-vanishing
value could be obtained by introducing an asymmetry of the
set-up through unequal wall-tensions which we will however not
do here. A further integration of $Y$ yields the warp-function
\begin{equation}
  A(x^5)=-2\ln\left|\sinh\left(\frac{k}{4}
                                \left[|x^5+l|+|x^5-l|-2a\right]
                          \right)
              \right| + b \; ,
  \label{WarpFunction}
\end{equation}
where $b$ is a second integration constant. Note, that the
above solution is valid for the parameter-range $T\ge 3M^3 k$
as can be easily recognized from (\ref{IntConstTension}). If $T
< 3M^3 k$, we have to substitute a ``tanh'' for the ``coth''
appearing in (\ref{YSol}) and (\ref{IntConstTension}), while
(\ref{CosmStep}) remains the same. This
 amounts
to a change from ``sinh'' to ``cosh'' in (\ref{WarpFunction})
Since we assume a positive wall-tension $T>0$, the integration
constant $a$ is constrained through (\ref{IntConstTension})
over the whole parameter-region, $T>0$, $\Lambda_e \le 0$, by
the lower bound $a>l$.

We have two free integration constants $a$ and $b$. We will see
however below that both are related with each other by the
imposition that for the situation with finetuned parameters we
should get back a solution with flat 4-dimensional sections.
Therefore we will end up with just one free integration
constant. This remaining free parameter will be fixed in such a
way that the finetuning limit exactly reproduces
(\ref{FlatSolution}) without any further constant added to
$A(x^5)$. This constitutes one fine-tuning at the classical
level as $A(x^5)$ is determined by the Einstein equations only
up to an additive constant.

The explicit solution shows that the warp-factor $f=e^{-A(x^5)}$
vanishes at $x^5=\pm a$. If $Q<0$ (which will turn out to be the
$\text{AdS}_4$ case, whereas the $\text{dS}_4$ case is free of
singularities) this gives rise to a singular 5-dimensional
curvature at these points
\begin{equation}
  \lim_{x^5 \rightarrow \pm a} R(G)
  \rightarrow
    \frac{24\Theta(-Q)}{(|x^5|-a)^2}
    \; , \qquad
    Q = \frac{T-3M^3 k}{T+3M^3 k}
    \; ,
\end{equation}
where the Heaviside step-function is defined by
$\Theta(x)=0,x<0$ and $\Theta(x)=1, x>0$. Due to the vanishing
of the warp-factor at these points we expect a tremendous
red-shift in signals originating there. Indeed, let us conceive
a wave signal emitted with frequency $\nu_e$ at $x^5=\pm a$.
Then that wave will be observed in the interior region $x^5 \in
(-a,a)$ with frequency $\nu_o$ given by
\begin{equation}
  \frac{\nu_o}{\nu_e} = \sqrt{\frac{G_{11}(x^5=\pm a)}{G_{11}(|x^5|<a)}}
                      = 0 \; ,
\end{equation}
due to the vanishing of the warp-factor at $x^5=\pm a$. Hence,
an infinite redshift makes it impossible for the region
$|x^5|\ge a$ to communicate to
 our
world (at least via electromagnetic radiation). Therefore, we
should
 restrict
the $x^5$ integration region to the causally connected interval
$x^5\in (-a,a)$.

Since recently there has been a discussion in the literature
\cite{Gubser},\cite{FLLN1},\cite{FLLN2} about which singularities are
permissible and which have better to be avoided, it is
interesting to see the verdict on our singularities in the case
of $Q<0$. In
\cite{Gubser} it has been argued that in a gravitational system exhibiting a
4-dimensional flat solution together with bulk scalars, only
those singularities are allowed, which leave the scalar
potential bounded from above. In our case, where we do not have
any scalars, the role of the scalar potential is played by the
bulk cosmological constant $\Lambda_e$ (together with the
tension $T$ at the wall-positions), which is clearly bounded
from above.  If the criterion of
\cite{Gubser} generalizes to the case where the 4-dimensional metric
deviates slightly (since in the end $\Lambda_4$ turns out to be
exponentially small) from the flat case, we would conclude that
the singularities encountered above for $Q<0$ are of the
permissible type.

Furthermore, in \cite{FLLN2} a consistency condition has been
derived which should hold for the effective cosmological
constant obtained by integration over the causally connected
$x^5$-region. We will now demonstrate that this consistency
condition is a simple consequence of
(\ref{FourScalarEquiv1}),(\ref{FourScalarEquiv2}) and the
expression (\ref{DefinitionLambda}), which defines $\Lambda_4$.
Starting with (\ref{DefinitionLambda}) and employing
(\ref{FourScalarEquiv1}),(\ref{FourScalarEquiv2}) to eliminate
the
 derivatives
$[(\ln f)^\prime]^2$ and $f^{\prime\prime}$,
(\ref{DefinitionLambda})
 becomes
\begin{equation}
  \Lambda_4 = 2\langle \Lambda_4 \rangle
             -\frac{1}{3}\int_{-a}^a dx^5 f^2 \left( 2\Lambda(x^5)
             +T\delta(x^5+l)+T\delta(x^5-l)\right) \; .
\end{equation}
Noticing that $f^\prime f(x^5=\pm a)=0$, we use
(\ref{DeterminingLambda}) to obtain
\begin{alignat}{3}
  \Lambda_4 &= \frac{1}{3}\int_{-a}^a dx^5 f^2 \left( 2\Lambda(x^5)
              +T\delta(x^5+l)+T\delta(x^5-l)\right) \notag \\
            &= -\frac{1}{3}\int_{-a}^a dx^5 f^2
               \left( T_1^{\phantom{1}1}+T_5^{\phantom{5}5} \right) \; ,
  \label{NillesConsistency}
\end{alignat}
which is nothing but the consistency condition of \cite{FLLN2}.
Since our solution has been derived from
(\ref{Consistency}),(\ref{FourScalarEquiv2}) which are
equivalent to (\ref{FourScalarEquiv1}),(\ref{FourScalarEquiv2})
and we will furthermore
 only
require (\ref{DefinitionLambda}) to obtain $\Lambda_4$, we
conclude that the consistency condition
(\ref{NillesConsistency}) of \cite{FLLN2} should be satisfied
for our solution.

After this short intermezzo on singularities, let us proceed by
inverting (\ref{IntConstTension}), to express $a$ explicitly
through the input values $T$ and $\Lambda_e$
\begin{equation}
   a = -\frac{1}{k}\ln |Q| + l \; ,
   \label{aConstant}
\end{equation}
which is valid for both $T \ge 3M^3 k$ and $T < 3M^3 k$. This
shows how the parameters $T,M,\Lambda_e$ influence the width of
the $x^5$ domain.

In order to determine $\Lambda_4(x^5)$, note that to obey the
Einstein equations, we have to fulfill
(\ref{FourScalarEquiv2}). This equation can be used to derive
the following expression for $\Lambda_4(x^5)$
\begin{equation}
            \Lambda_4(x^5)
          = \pm\frac{3}{2}e^{-b}M^2_{\text{eff}}
            \left\{ \begin{array}{cc}
                    k^2   &,\;\; |x^5| > l \\
                    k^2/4 &,\;\; x^5 = \pm l \\
                      0      &,\;\; |x^5| < l
                    \end{array}
            \right.
     \label{SectionCoCo}
\end{equation}
where the plus-sign applies to the case $T\ge 3M^3 k$, whereas
the minus-sign applies to the complementary case in which $T <
3M^3 k$. Since we do not want to use $\Lambda_4(x^5)$ as an
input to determine $b$, but rather focus on the opposite, we
are looking for an additional constraint, which allows for a
determination of the constant $b$. This extra constraint comes
from considering a smooth transition to the flat solution
(\ref{FlatSolution}) with $\Lambda_4=0$. As can be seen from
(\ref{Tension}), we reach the flat limit by sending
$T\rightarrow 3M^3 k$. Via (\ref{aConstant}) this limit
corresponds to sending the constant $a\rightarrow \infty$. Thus
we see, that the integration region $x^5 \in (-a,a)$ extends
over the whole real line in this limit and the warp-function
(\ref{WarpFunction}) becomes
\begin{equation}
  A(x^5) \rightarrow \frac{k}{2}\left(|x^5+l|+|x^5-l|\right)
                     +2\ln 2-ka+b \; .
\end{equation}
Thus, to guarantee a smooth transition to the flat solution
(\ref{FlatSolution}), we have to identify the integration
constants $a$ and $b$ as follows
\begin{equation}
  b = -2\ln 2+ka \; .
\end{equation}

Notice that here we implicitly used the mentioned finetuning as
$A(x^5)$ in (\ref{FlatSolution}) is only determined by the D=5
Einstein equations up to an additive constant which we have set
to zero. Thus, together with (\ref{aConstant}) and
(\ref{SectionCoCo}) we obtain the following expression for
$\Lambda_4(x^5)$
\begin{equation}
     \Lambda_4(x^5)
   =  6 e^{-kl} Q M^2_{\text{eff}}
     \left\{ \begin{array}{cc}
                k^2     &,\; |x^5| > l \\
                k^2/4   &,\; x^5 = \pm l \\
                   0        &,\; |x^5| < l
             \end{array}
     \right.
    \label{SectionCoCoFinal}
\end{equation}
which is valid for both parameter-regions $T \ge 3M^3 k$ and $T
< 3M^3 k$.

Finally, to obtain the effective four-dimensional $\Lambda_4$,
we have to take the mean of $\Lambda_4(x^5)$. Again using that
$f^\prime f(x^5=\pm a)=0$, we employ (\ref{DeterminingLambda})
and arrive at
\begin{equation}
  \Lambda_4 = \frac{\int_{-a}^a dx^5 e^{-A(x^5)}\Lambda_4(x^5)}
                    {\int_{-a}^a dx^5 e^{-A(x^5)}}
            = 12e^{-2kl}M^3 k Q F(|Q|) \; ,
  \label{PreLambda}
\end{equation}
where we defined $F(|Q|)=1-|Q|^2+2|Q|\ln|Q|$. In addition we
obtain the following effective Planck-scale
\begin{equation}
  M_{\text{eff}}^2 = M^3 \int_{-a}^a dx^5 e^{-A(x^5)}
                   = 2e^{-kl}M^3\left(l(1-|Q|)^2+\frac{F(|Q|)}{k}\right)
                      \; .
  \label{PreMass}
\end{equation}

There is an exponential-factor occuring in $\Lambda_4$ which is
the square of the one occuring in $M_{\text{eff}}^2$. At the
classical level (classical in the bulk of the five-dimensional
spacetime -- the field-theories on the walls are however
considered quantum mechanically!) an overall constant $e^{-kl}$
multiplying the whole effective 4-dimensional action
\begin{alignat}{3}
 S_{D=4} &=
-\int d^4x\sqrt{-g}(M^2_{\text{eff}}R(g)+\Lambda_4) \\ \notag
&= -e^{-kl}\int d^4x\sqrt{-g}({\tilde
M}^2_{\text{eff}}R(g)+{\tilde
\Lambda}_4)  \\ \notag
&= -e^{-kl}{\tilde M}^2_{\text{eff}}\int
d^4x\sqrt{-g}(R(g)+\lambda_4)
\end{alignat}
is immaterial -- it simply drops out of the field
equation\footnote{Actually, by starting with a more general
Ansatz for the five-dimensional metric in which $(dx^5)^2$ gets
replaced by $e^{B(x^5)}(dx^5)^2$ avoids this overall
multiplicative constant at all \cite{KU}.}. Therefore, we can
neglect the overall factor $e^{-kl}$. The physically observable
cosmological constant -- invariant under any overall rescaling
-- is given by $\lambda_4=\Lambda_4/M^2_{\text{eff}}= {\tilde
\Lambda_4}/{\tilde M}^2_{\text{eff}}$. With (\ref{PreLambda}) and
(\ref{PreMass}) we thus obtain our final result
\begin{equation}
  \lambda_4 = e^{-kl}
              \left( \frac{6k^2 Q F(|Q|)}{kl(1-|Q|)^2 + F(|Q|)}
              \right) \; .
\end{equation}

Some comments about this formula are in order. First, the
physical range of the parameter $Q$ lies between $0\le |Q|\le
1$, where we presuppose a non-negative wall-tension $T>0$. The
lower bound corresponds to the finetuned flat $\Lambda_4=0$
limit, while the upper bound is reached for vanishing bulk
cosmological constant $\Lambda_e=0$. Over that region we have $1
\ge F(|Q|<1)>0$ and $F(1)=0$. Hence, we see that starting with
some given values for $\Lambda_e\le 0,M,T>0$ we obtain a
positive or negative $\lambda_4$ depending on the sign of $Q$.
For $T>\sqrt{-3M^3\Lambda_e}$ the 4-dimensional spacetime will
be $\text{dS}_4$, whereas for $T<\sqrt{-3M^3\Lambda_e}$ it will
be $\text{AdS}_4$. Furthermore, we recognise a smooth
connection to the case with flat 4-dimensional Minkowski
spacetime for finetuned parameters $T=\sqrt{-3M^3\Lambda_e}
\Leftrightarrow Q=0$. Second, there is no need for a finetuning
of the fundamental parameters to receive a small $\lambda_4$.
By adapting the distance $2l$ between both walls, one arrives
at a huge enough suppression through the exponential factor
such that the observed value could be accounted for. Moreover,
thanks to the exponential suppression this does not amount to
an extremely large hierarchy between the fundamental scale $M$
and the separation-scale $1/2l$.

Hence at the price of one classical finetuning (the additive
integration constant for $A(x^5)$ had been set to zero) one is
able to bring the generically Planck-sized contribution to
$\lambda_4$ coming from the three fundamental parameters
$M,T,\Lambda_e$ down to a small value which for $2l=1/M_{GUT}$
agrees with the experimental observation that
$\lambda_4\simeq(\text{meV})^4/M_{Pl}^2$ \cite{AK3}. It is
important to realise that these three parameters contain all
the quantum corrections coming from matter fields (including
the Standard Model ones) which are located on the walls.
Therefore one does not require any more an infinite finetuning
order by order in loop corrections as in the case of a
four-dimensional field theory coupled to four-dimensional
gravity to match the Cosmological Constant with its observed
value. Of course one of the interesting problems which remains
is the stabilisation of the inter-wall distance. Without gauge
fields in the bulk, mechanisms like the one proposed in
\cite{LS} are not applicable and it would probably be most
satisfying to embed the five-dimensional brane-world first into
string- or M-theory and then to study the forces between the
branes directly in M-theory \cite{K1} and to incorporate generic
non-perturbative M-theory effects like open membrane instantons
for its stabilisation \cite{CK2}. Alternatively one might try
to use a Goldberger-Wise like mechanism \cite{GW} for the
stabilization directly in five dimensions.

\bigskip
\noindent {\large \bf Acknowledgements}\\[2ex]
The author would like to thank the DFG and the Graduiertenkolleg
``Theoretische Elementarteilchenphysik'' for financial support.

 \newcommand{\zpc}[3]{{\sl Z. Phys.} {\bf C\,#1} (#2) #3}
 \newcommand{\npb}[3]{{\sl Nucl. Phys.} {\bf B\,#1} (#2)~#3}
 \newcommand{\plb}[3]{{\sl Phys. Lett.} {\bf B\,#1} (#2) #3}
 \newcommand{\prd}[3]{{\sl Phys. Rev.} {\bf D\,#1} (#2) #3}
 \newcommand{\prb}[3]{{\sl Phys. Rev.} {\bf B\,#1} (#2) #3}
 \newcommand{\pr}[3]{{\sl Phys. Rev.} {\bf #1} (#2) #3}
 \newcommand{\prl}[3]{{\sl Phys. Rev. Lett.} {\bf #1} (#2) #3}
 \newcommand{\jhep}[3]{{\sl JHEP} {\bf #1} (#2) #3}
 \newcommand{\cqg}[3]{{\sl Class. Quant. Grav.} {\bf #1} (#2) #3}
 \newcommand{\atmp}[3]{{\sl Adv. Theor. Math. Phys.} {\bf #1} (#2) #3}
 \newcommand{\prep}[3]{{\sl Phys. Rep.} {\bf #1} (#2) #3}
 \newcommand{\fp}[3]{{\sl Fortschr. Phys.} {\bf #1} (#2) #3}
 \newcommand{\nc}[3]{{\sl Nuovo Cimento} {\bf #1} (#2) #3}
 \newcommand{\nca}[3]{{\sl Nuovo Cimento} {\bf A\,#1} (#2) #3}
 \newcommand{\lnc}[3]{{\sl Lett. Nuovo Cimento} {\bf #1} (#2) #3}
 \newcommand{\ijmpa}[3]{{\sl Int. J. Mod. Phys.} {\bf A\,#1} (#2) #3}
 \newcommand{\rmp}[3]{{\sl Rev. Mod. Phys.} {\bf #1} (#2) #3}
 \newcommand{\ptp}[3]{{\sl Prog. Theor. Phys.} {\bf #1} (#2) #3}
 \newcommand{\sjnp}[3]{{\sl Sov. J. Nucl. Phys.} {\bf #1} (#2) #3}
 \newcommand{\sjpn}[3]{{\sl Sov. J. Particles \& Nuclei} {\bf #1} (#2) #3}
 \newcommand{\splir}[3]{{\sl Sov. Phys. Leb. Inst. Rep.} {\bf #1} (#2) #3}
 \newcommand{\tmf}[3]{{\sl Teor. Mat. Fiz.} {\bf #1} (#2) #3}
 \newcommand{\jcp}[3]{{\sl J. Comp. Phys.} {\bf #1} (#2) #3}
 \newcommand{\cpc}[3]{{\sl Comp. Phys. Commun.} {\bf #1} (#2) #3}
 \newcommand{\mpla}[3]{{\sl Mod. Phys. Lett.} {\bf A\,#1} (#2) #3}
 \newcommand{\cmp}[3]{{\sl Comm. Math. Phys.} {\bf #1} (#2) #3}
 \newcommand{\jmp}[3]{{\sl J. Math. Phys.} {\bf #1} (#2) #3}
 \newcommand{\pa}[3]{{\sl Physica} {\bf A\,#1} (#2) #3}
 \newcommand{\nim}[3]{{\sl Nucl. Instr. Meth.} {\bf #1} (#2) #3}
 \newcommand{\el}[3]{{\sl Europhysics Letters} {\bf #1} (#2) #3}
 \newcommand{\aop}[3]{{\sl Ann. of Phys.} {\bf #1} (#2) #3}
 \newcommand{\jetp}[3]{{\sl JETP} {\bf #1} (#2) #3}
 \newcommand{\jetpl}[3]{{\sl JETP Lett.} {\bf #1} (#2) #3}
 \newcommand{\acpp}[3]{{\sl Acta Physica Polonica} {\bf #1} (#2) #3}
 \newcommand{\sci}[3]{{\sl Science} {\bf #1} (#2) #3}
 \newcommand{\vj}[4]{{\sl #1~}{\bf #2} (#3) #4}
 \newcommand{\ej}[3]{{\bf #1} (#2) #3}
 \newcommand{\vjs}[2]{{\sl #1~}{\bf #2}}
 \newcommand{\hepph}[1]{{\sl hep--ph/}{#1}}
 \newcommand{\desy}[1]{{\sl DESY-Report~}{#1}}

\bibliographystyle{plain}

\end{document}